\documentclass[12pt]{article}
\usepackage{docmute}

\pdfoutput=1
\usepackage{epsf}
\usepackage{amsmath,amssymb}
\usepackage{pdfpages}
\usepackage{graphicx}
\usepackage{comment}
\usepackage{tikz}
\usetikzlibrary{positioning}
\usepackage{physics}
\usepackage{caption}
\usepackage{subcaption}
\usepackage{float}
\usepackage[sort&compress, numbers, merge]{natbib}

\begin{document}

\renewcommand{\thefootnote}{\fnsymbol{footnote}}
\setcounter{footnote}{0}

\begin{titlepage}

\def\thefootnote{\fnsymbol{footnote}}

\begin{center}
\hfill June, 2022\\
\hfill TU-1159\\
\vskip .25in

{\Large \bf

  Instability of the Electroweak Vacuum 
  \\[2mm]
  in Starobinsky Inflation 

}

\vskip .5in

{\large
  Qiang Li$^{(a)}$, Takeo Moroi$^{(a)}$, Kazunori Nakayama$^{(b,c)}$
  and Wen Yin$^{(b)}$
}

\vskip 0.5in

$^{(a)}${\em
  Department of Physics, The University of Tokyo, Tokyo 113-0033, Japan
}

\vskip 0.2in

$^{(b)}${\em 
Department of Physics, Tohoku University, Sendai 980-8578, Japan
}

\vskip 0.2in

$^{(c)}${\em 
International Center for Quantum-field Measurement Systems for Studies of the Universe and Particles (QUP), KEK, 1-1 Oho, Tsukuba, Ibaraki 305-0801, Japan
}

\end{center}
\vskip .5in

\begin{abstract}

  We study the stability of the electroweak vacuum during and after
  the Starobinsky inflation, assuming the existence of the non-minimal
  Higgs coupling to the Ricci scalar.  In the Starobinsky inflation,
  there exists $R^2$ term (with $R$ being the Ricci scalar), which
  modifies the evolution equation of the Higgs field.  We consider the
  case that the non-minimal coupling is sizable so that the quantum
  fluctuation of the Higgs field is suppressed and that the Higgs
  amplitude is settled near the origin during the inflation.  In such
  a case, the Higgs amplitude may be amplified in the preheating epoch
  after inflation because of the parametric resonance due to the
  non-minimal coupling.  We perform a detailed analysis of the
  evolution of the Higgs field in the preheating epoch by a numerical
  lattice simulation and derive an upper bound on the non-minimal
  coupling constant $\xi$ in order to realize the electroweak vacuum
  in the present universe.  We find that the upper bound on $\xi$ in
  the Starobinsky inflation model is more stringent than that in
  conventional inflation models without the $R^2$ term.

\end{abstract}

\end{titlepage}

\renewcommand{\thepage}{\arabic{page}}
\setcounter{page}{1}
\renewcommand{\thefootnote}{\#\arabic{footnote}}
\setcounter{footnote}{0}
\renewcommand{\theequation}{\thesection.\arabic{equation}}

\section{Introduction}
\label{sec:intro}
\setcounter{equation}{0}

Stability of the electroweak vacuum, in which we are living, is highly
non-trivial in quantum field theory.  Even if the electroweak vacuum,
at which the Higgs vacuum expectation value (VEV) is given by $\langle
h\rangle\simeq 246\ {\rm GeV}$, corresponds to a local minimum of the
Higgs potential, there may exist another minimum of the potential at
which the energy density becomes smaller than that of the electroweak
vacuum.  If so, the electroweak vacuum becomes metastable and it
decays into the true vacuum via the quantum tunneling
effect~\cite{Sher:1988mj,Arnold:1989cb,Anderson:1990aa,Arnold:1991cv,Espinosa:1995se,Isidori:2001bm,Espinosa:2007qp,Ellis:2009tp}.
The metastability of the electroweak vacuum occurs in the standard
model as well as in certain models with physics beyond the standard
model.

In the standard model, it is well known that the Higgs quartic
coupling, which is positive at the electroweak scale, may become
negative at higher energy scale due to the renormalization group
effect.  Using the central values of standard-model parameters, the
Higgs quartic coupling constant becomes negative at the instability
scale of $\sim O(10^{10})\ {\rm GeV}$.  The negativity of the quartic
coupling constant indicates that the electroweak vacuum is not the
absolute minimum of the potential and that it is metastable.  We
emphasize that the metastability of the electroweak vacuum does not
imply the difficulty to realize the electroweak vacuum in the present
universe.  Indeed, in the standard model, the lifetime of the
electroweak vacuum is much longer than the present cosmic time
\cite{EliasMiro:2011aa,Bezrukov:2012sa,Degrassi:2012ry, Buttazzo:2013uya,Bednyakov:2015sca,Andreassen:2017rzq, Chigusa:2017dux,
  Chigusa:2018uuj}.  Thus, once the Higgs field settles to the
electroweak vacuum in the early universe, we can safely live in the
electroweak vacuum even if the standard model is valid up to a very
high energy scale.

The behavior of the Higgs field is, however, highly non-trivial in the
early universe.  In particular, during and after the inflation, the
Higgs field is influenced by the dynamics of the rapid expansion of
the universe as well as by the motion of the inflaton.  During the
inflation, the quantum fluctuation of the Higgs field may make the
Higgs amplitude larger than the instability scale; in such a case, the
Higgs shows a run-away behavior during inflation due to the negative
quartic coupling, which provides a cosmic history inconsistent with
the present universe~\cite{Kobakhidze:2013tn,Fairbairn:2014zia,Hook:2014uia,Herranen:2014cua,Kamada:2014ufa,Kearney:2015vba,Espinosa:2015qea,Mantziris:2020rzh}.  
Such a problem can be avoided if the Higgs
field has a non-minimal coupling to the Ricci scalar. The non-minimal
coupling induces an effective mass term of the Higgs during the
inflation which stabilizes the Higgs potential if the sign of the
non-minimal coupling constant is properly chosen.  Hereafter, we
concentrate on the case with the non-minimal coupling of the Higgs.
Even though the non-minimal coupling stabilizes the Higgs potential
during the inflation, it may cause an instability after the inflation~\cite{Herranen:2015ima,Ema:2016kpf,Kohri:2016wof,Enqvist:2016mqj,Postma:2017hbk,Ema:2017loe,Ema:2017rkk,Figueroa:2017slm,Rusak:2018kel,Kost:2021rbi}.
With the non-minimal coupling, the effective mass of the Higgs may
have significant time-dependence because of the oscillatory behavior
of the Ricci scalar after inflation.  Then, the Higgs amplitude may be
amplified due to the parametric resonance~\cite{Dolgov:1989us,Traschen:1990sw,Kofman:1994rk,Shtanov:1994ce,Kofman:1997yn} or tachyonic resonance~\cite{Bassett:1997az,Tsujikawa:1999jh,Dufaux:2006ee} at the preheating epoch
after the inflation; if the effect of the parametric resonance is too
large, the Higgs amplitude exceeds the instability scale and the Higgs
shows the run-away behavior.  The effect of the parametric resonance
is more enhanced with larger value of the non-minimal coupling, and we
obtain an upper bound on the non-minimal coupling to realize the
electroweak vacuum in the present universe.

The dynamics of the Higgs field after inflation depends on couplings
of the Higgs to the inflaton and Ricci scalar as well as on the model
of the inflation.  The upper bound on the non-minimal coupling has
been studied in simple inflation model in which the gravity sector is
described by the Einstein-Hilbert action~\cite{Herranen:2015ima,Ema:2016kpf,Kohri:2016wof,Enqvist:2016mqj,Postma:2017hbk,Ema:2017loe,Ema:2017rkk,Figueroa:2017slm,Rusak:2018kel,Kost:2021rbi}. Based on the
recent observations of cosmic density perturbations, however, an
inflation model with $R^2$ term (with $R$ being the Ricci scalar),
which is called the Starobinsky inflation \cite{Starobinsky:1980te}, has been
attracted many attentions.  The Starobinsky inflation predicts the
scalar spectral index and the tensor-to-scalar ratio consistent with
the observations \cite{Akrami:2018odb}.  In addition, the Starobinsky inflation
provides an interesting possibility of producing hidden-sector dark matter
via the decay of the inflaton \cite{Gorbunov:2010bn,Gorbunov:2012ij,Bernal:2020qyu,Li:2021fao}.  Phenomenology based on the
Starobinsky inflation crucially depends on the stability of the
electroweak vacuum during and after the inflation.  Importantly, the
evolution equation of the Higgs field in Starobinsky inflation differs
from that in simple inflation models (without the $R^2$ term).  Thus,
the dedicated study about the stability of the electroweak vacuum is
necessary for the case of the Starobinsky inflation.

In this paper, we consider the stability of the electroweak vacuum
during and after the Starobinsky inflation. In particular, we study in
detail the Higgs dynamics after inflation by a numerical lattice
simulation.  Then, we derive an upper bound on the non-minimal
coupling constant to realize the electroweak vacuum in the present
universe.

The organization of this paper is as follows.  In Section
\ref{sec:model}, we give an overview of the Starobinsky inflation
model as well as the behavior of the Higgs and inflaton potential in
the framework of our interest.  In Section \ref{sec:instability}, we
discuss the stability of the electroweak vacuum during and after the
Starobinsky inflation.  In Section \ref{sec:lattice}, we perform a
lattice simulation to study the stability of the electroweak vacuum in
the preheating epoch after inflation and derive an upper bound on the
non-minimal coupling constant.  Section \ref{sec:conclusions} is
devoted to conclusions and discussion.

\section{Model}
\label{sec:model}
\setcounter{equation}{0}

In this section, we summarize the basic features of the Starobinsky
inflation model with the Higgs non-minimal
coupling to gravity.  We also give a brief summary of the properties
of the Higgs potential in the standard model.

\subsection{Lagrangian}

We start with introducing the total Lagrangian of the model we
consider.  In the Jordan frame, the action has the following form:
\begin{align}
  \label{action}
  S=S_{\rm{inf}}+S_{\rm{Higgs}}+S_{\rm{int}},
\end{align}
where $S_{\rm{inf}}$, $S_{\rm{Higgs}}$, $S_{\rm{int}}$ are actions of
the inflation sector, the Higgs sector and the interaction between
Higgs and gravity, respectively.  Taking the unitary gauge
$\Phi=(0,h/\sqrt{2})^T$ (with $\Phi$ being the Higgs doublet while $h$
being a real scalar field), they are given by
\cite{Starobinsky:1980te}
\begin{align}
  S_{\rm{inf}} &= \int d^4x\sqrt{-\hat{g}}\left[-\frac{M_{\rm Pl}^2}{2}\left(\hat{R}-\frac{1}{6\mu^2}\hat{R}^2\right)\right],\\
  S_{\rm{Higgs}} &= \int d^4x\sqrt{-\hat{g}}\left(\frac{1}{2}\hat{g}^{\mu\nu}\partial_\mu h\partial_\nu h-\frac{\lambda}{4}h^4\right),\\
  S_{\rm{int}} &= \int d^4x\sqrt{-\hat{g}}\ \frac{1}{2}\xi \hat{R}h^2,
  \label{int}
\end{align}
where fields with hat are defined in the Jordan frame, $M_{\rm Pl}\simeq
2.4\times 10^{18}\ {\rm GeV}$ is the reduced Planck scale, and $\xi$ is
the Higgs-gravity non-minimal coupling constant.\footnote
{We neglect the bare Higgs quadratic term which is irrelevant for our
  discussion.}
Hereafter, we consider the case of $\xi$ being non-negative.\footnote
{In our convention, the conformal coupling is $\xi=1/6$. If $\xi<0$,
  the non-minimal coupling induces a tachyonic mass term of the Higgs
  field, so the electroweak vacuum is destabilized during inflation
  \cite{Espinosa:2015qea}.}
The Higgs quartic coupling constant $\lambda$ depends on the
renormalization scale $Q$.  More detail about the scale dependence of
$\lambda$ will be discussed in the next subsection.

By introducing an auxiliary field $\varphi$ \cite{Maeda:1988ab}, the action (\ref{action}) can be rewritten as 
\begin{align}\label{aux}
S=\int d^4x \sqrt{-\hat{g}}\left[-\frac{M_{\rm Pl}^2}{2}\left(1-\frac{\varphi}{3\mu^2}-\frac{\xi}{M_{\rm Pl}^2}h^2\right)\hat{R}-\frac{M_{\rm Pl}^2\varphi^2}{12\mu^2}+\frac{1}{2}\hat{g}^{\mu\nu}\partial_\mu h\partial_\nu h-\frac{\lambda}{4}h^4\right].
\end{align}
Note that the Euler-Lagrange equation of $\varphi$ gives
\begin{align}
\varphi=\hat{R},
\end{align}
and by substituting back to (\ref{aux}) we obtain the original action.

For the study of the stability of the electroweak vacuum, it is
convenient to work in the Einstein frame. With the conformal
transformation of the metric
\begin{align}
  g_{\mu\nu}=\Omega^2\hat{g}_{\mu\nu},
\end{align}
where
\begin{align}
  \Omega^2=1-\frac{\varphi}{3\mu^2}-\frac{\xi}{M_{\rm Pl}^2}h^2,
\end{align}
we can eliminate the non-minimal couplings and obtain the action in
the Einstein frame as
\begin{align}\label{einac}
S=\int d^4x \sqrt{-g}\left[-\frac{M_{\rm Pl}^2}{2}R+\frac{1}{2}g^{\mu\nu}\partial_\mu\phi\partial_\nu\phi+\frac{1}{2}e^{-\chi}g^{\mu\nu}\partial_\mu h\partial_\nu h-V(\phi,h)\right],
\end{align}
where we have defined $\chi$ and the scalaron field $\phi$ through
\begin{align}
  \frac{\phi}{M_{\rm Pl}}=\sqrt{\frac{3}{2}}\ln\Omega^2,~~~
  \chi=\sqrt{\frac{2}{3}}\frac{\phi}{M_{\rm Pl}}.
\end{align}
Although $\phi$ (or, equivalently, $\varphi$) was introduced as the
auxiliary field, it becomes a physical degree of freedom; with the
$R^2$ term, there exists an extra physical degree of freedom in the
metric other than the tensor modes, and it is converted to $\phi$ by
the conformal transformation.  In addition, the scalar potential
$V(\phi,h)$ is given by
\begin{align}\label{potential}
  V(\phi,h) = 
  \frac{3\mu^2M_{\rm Pl}^2}{4}
  \left(1-e^{-\chi}+\frac{\xi}{M_{\rm Pl}^2}e^{-\chi}h^2\right)^2
  + \frac{\lambda}{4} e^{-2\chi} h^4.
\end{align}

If the initial amplitude of $\phi$ is larger than $M_{\rm Pl}$, an
approximate de Sitter space is realized and the inflation occurs.
This can be understood by studying the potential of $\phi$ (with
neglecting the Higgs field):
\begin{align}
  V\simeq \frac{3\mu^2M_{\rm Pl}^2}{4}\left[1-\exp\left(-\sqrt{\frac{2}{3}}\frac{\phi}{M_{\rm Pl}}\right)\right]^2,
\end{align}
One can see that the potential becomes flat when $\phi\gg M_{\rm Pl}$.
Due to the flatness of this potential at large $\phi$, the slow-roll
inflation (called Starobinsky inflation) can happen.  
The expansion rate during the inflation is evaluated as
\begin{align}
  H_{\rm inf} \simeq \frac{1}{2} \mu.
\end{align}
We define $t_*$ as the time when the slow roll parameter,
$\epsilon\equiv -\dot{H}/H^2$, becomes equal to unity (i.e., end of
the inflation); at the time of $t=t_*$,
\begin{align}
  \epsilon (t_*) = 1,~~~
  \phi_* \equiv \phi (t_*) \simeq 0.97 M_{\rm Pl},~~~
  \dot{\phi}_* \equiv \dot{\phi} (t_*) \simeq 
  -3.75 \times 10^{-6} M_{\rm Pl}^2.
  \label{t*}
\end{align}

In the Starobinsky inflation model, the curvature perturbation
amplitude $A_s$, the scalar spectral index $n_s$, and the
tensor-to-scalar ratio $r$ are evaluated as
\begin{align}
A_s(k) &\simeq\frac{1}{24\pi^2}\frac{\mu^2}{M_{\rm Pl}^2}N_e^2,\\
n_s(k)-1 &\simeq -\frac{2}{N_e},\\
r(k) &\simeq\frac{12}{N_e^2},
\end{align}
where $N_e$ is the $e$-folding number at which the mode with comoving
wavenumber $k$ exits the horizon and is related to the scalaron amplitude as
\begin{align}
  N_e \simeq \frac{3}{4} \exp
  \left( \sqrt{\frac{2}{3}} \frac{\phi}{M_{\rm Pl}} \right).
\end{align}
Taking $N_e(k=0.05\rm{Mpc}^{-1})\simeq 56$, the observed value of
$A_s\simeq 2.1\times 10^{-9}$ \cite{Aghanim:2018eyx} gives
\begin{align}
  \mu\simeq 3.1\times 10^{13}\ {\rm GeV},
\end{align}
which will be used for our numerical analysis, while the scalar
spectral index and the tensor-to-scalar ratio are well within the
allowed region \cite{Akrami:2018odb,BICEP:2021xfz}.

We also note here that, during the inflation, the Higgs field acquires
an effective mass squared of $\sim 12\xi H_{\rm{inf}}^2$.  Thus, if
$\xi$ is larger than $\sim 0.1$, the Higgs field is forced to be at
the origin during the inflation.

\subsection{Higgs potential at quantum level}

Next, we discuss the behavior of the Higgs potential with including
the quantum effects.  The Higgs potential is dominated by the quartic
term as indicated in the previous section.  The coupling constant
$\lambda$ for the quartic interaction of the Higgs has scale
dependence and, as is well known, it may become negative at the scale
much higher than the electroweak scale.  In order to take account of
the scale dependence of the quartic coupling constant, we evaluate
$\lambda$ at the scale of the Higgs amplitude.  Because we
will deal with the case that the Higgs field is inhomogeneous, we
approximate the Higgs potential as
\begin{align}
  V_{\rm Higgs} = \frac{1}{4} \lambda (Q=\sqrt{\langle h^2\rangle} ) h^4.
\end{align}
where $\lambda (Q)$ denotes the quartic coupling constant at the
renormalization scale $Q$ and $\langle\cdots\rangle$ is the spatial
average.  The bare mass term of the Higgs is neglected because it is
irrelevant for our following discussion.

In our analysis, we assume the particle content of the standard model
(as well as inflaton) and study the scale dependence of $\lambda$.
The renormalization group behavior of $\lambda$ is sensitive to
standard model parameters, in particular, the top quark mass $M_t$,
the strong coupling constant $\alpha_s$, and the Higgs mass.

Let us first consider the top quark pole mass $M_t$.  It can be
obtained from the kinematics in the top anti-top events.  The latest
PDG average gives \cite{ParticleDataGroup:2020ssz}
\begin{align}
  \label{direcMt}
  M_t=172.76 \pm 0.3 \,{\rm GeV}.
\end{align}
We take the central value of $M_t$ as our canonical value and, in
order to take account of the top-mass uncertainty, we also provide the
numerical results with several values of the top mass. For the strong
coupling constant, we adopt \cite{ParticleDataGroup:2020ssz}
\begin{align}\label{as}
  \alpha_s (m_Z) =0.1179(9).
\end{align}
In addition, the Higgs boson mass is given by
\cite{ParticleDataGroup:2020ssz}
\begin{align}
  m_h=125.25\pm 0.17\ {\rm GeV}.
\end{align}

\begin{figure}[t]
  \centering
  \includegraphics[width=0.45\textwidth]{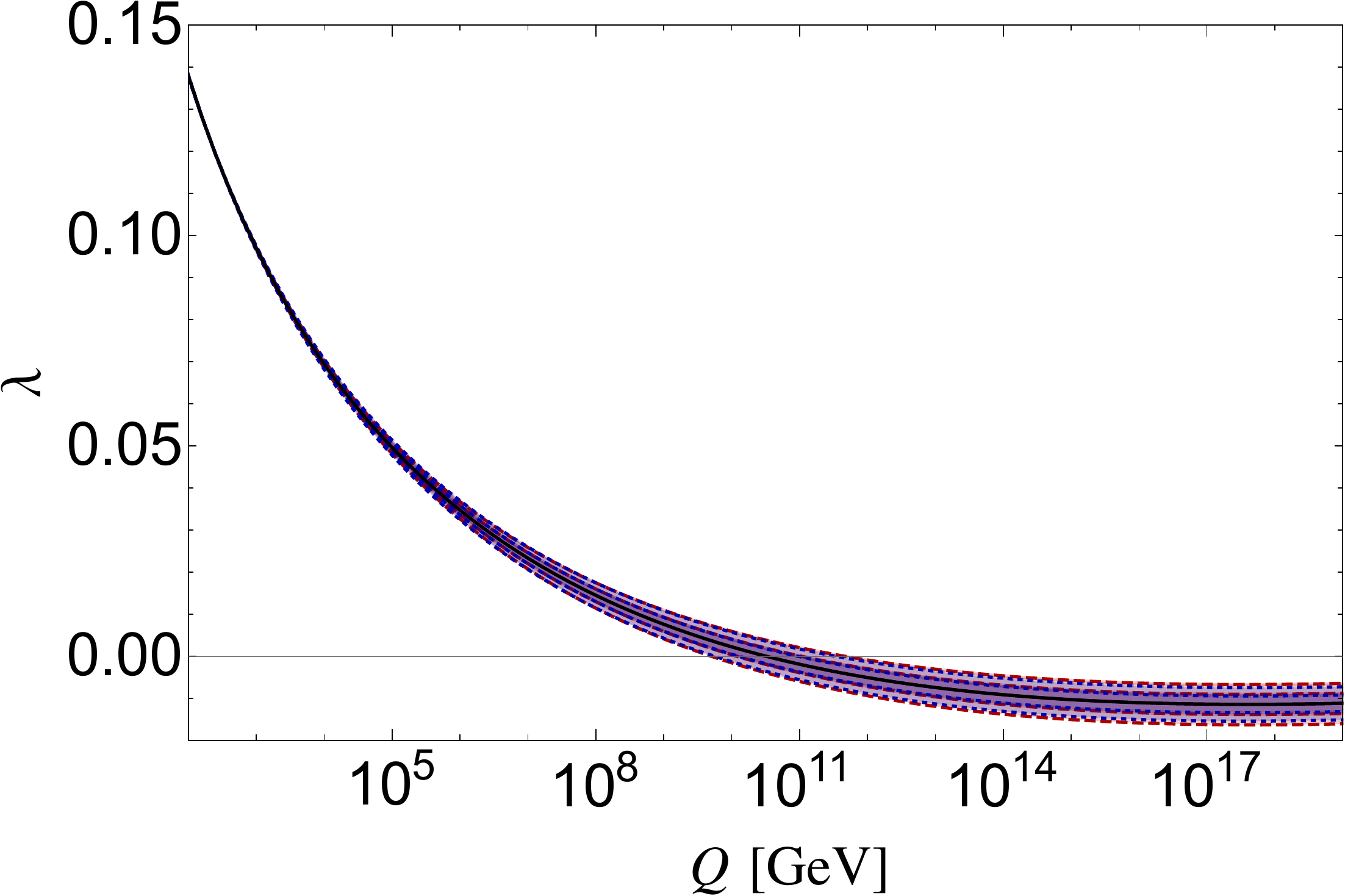}
  \includegraphics[width=0.45\textwidth]{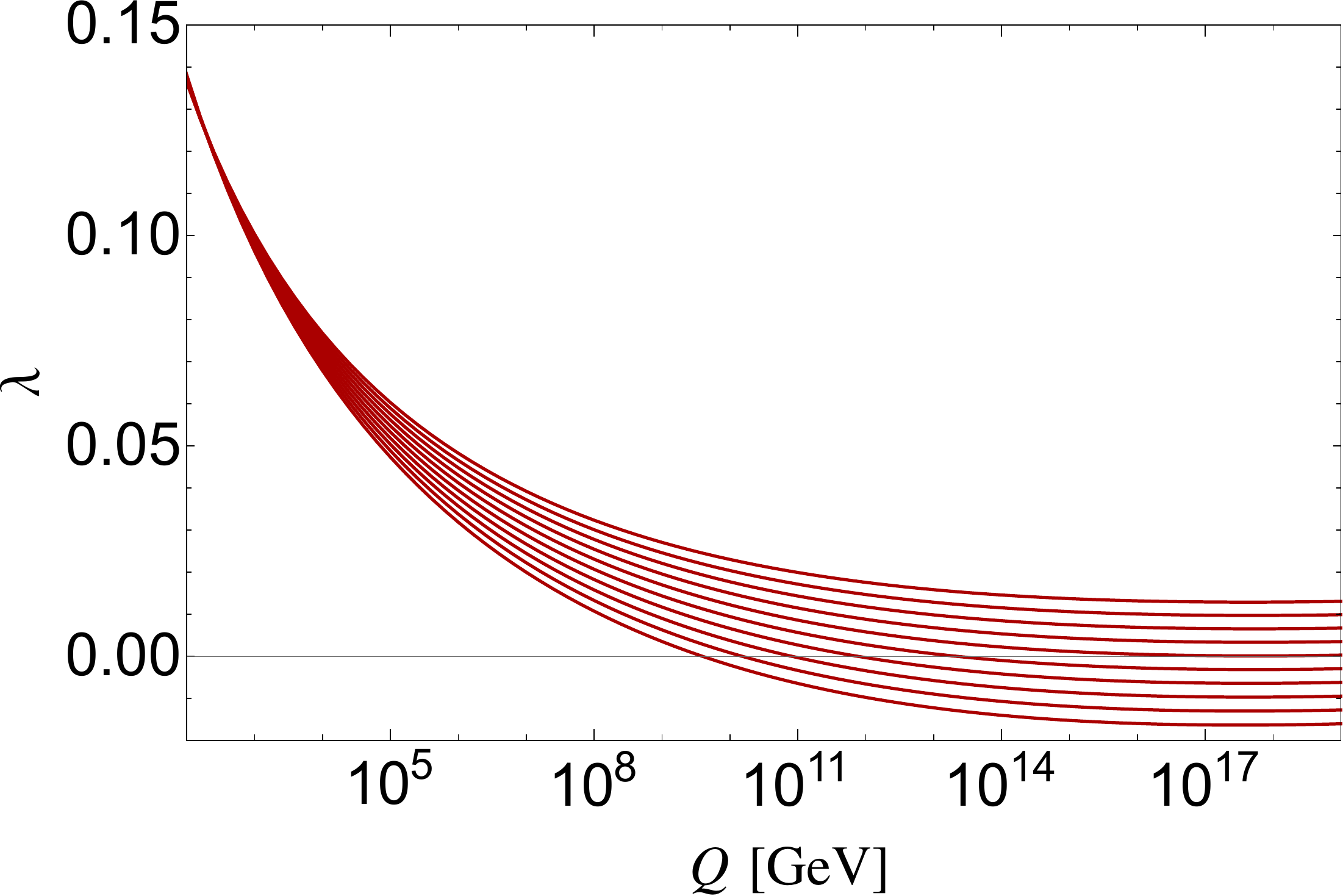}
  \caption{The Higgs quartic coupling $\lambda$ as a function of the
    renormalization scale $\mu_{\rm RG}$ with different input
    standard-model couplings.  In the left panel we show the running
    $\lambda$ with the central values of the measured standard-model
    couplings including \eqref{direcMt} (black solid line).  The blue
    dotted (red dashed) lines indicate the 1 and 2 $\sigma$ range by
    varying $M_t$ ($\alpha_s$) according to Eq.\ \eqref{as}
    (Eq.\ \eqref{direcMt}).  In the right panel, we show the scale
    dependence of $\lambda$ with $M_t=169.0$, $169.5$, $170.0$,
    $\cdots$, $173.5$ GeV from top to bottom.}
  \label{fig:lambda}
\end{figure}

We show the scale dependence of $\lambda$ in Fig.\ \ref{fig:lambda}.
We use the {\tt SMDR} code \cite{Martin:2019lqd}, which partially
includes 3, 4, and 5 loop effects, to numerically solve the
renormalization group equations in the standard model.  In the left
panel, $M_t$ and $\alpha_s$ are varied within 2 $\sigma$ ranges around
their central values.  We can see that even adopting such
uncertainties in the standard model parameters, $\lambda$ becomes
negative at a high scale and the Higgs potential is metastable. Using
the central values of the standard model parameters, we find the
instability scale, define as $\lambda (\Lambda_I)=0$, to be
\begin{align}
  \Lambda_I\simeq 3.3 \times 10^{10}\,{\rm GeV}.
\end{align}
The instability scale may vary by an order of magnitude when we take
account of the $\sim 2\sigma$ uncertainties.  In the right panel, we
show the scale dependence for several values of $M_t$ while taking
central values of other parameters. The Higgs potential becomes
absolutely stable for $M_t \lesssim 171$GeV, which is inconsistent
with Eq.\,\eqref{direcMt} at $\sim 6\sigma$ level.  Thus in the
standard model, the Higgs potential is very likely to have a radiative
instability.

\subsection{Effective mass of the Higgs}

The Higgs dynamics becomes highly non-trivial due to the presence of
inflaton field.  In the flat space-time (i.e., when the inflaton is at
the minimum of its potential), the Higgs field can just stay at the
electroweak vacuum.  During and after the inflation, on the contrary,
the inflaton is in motion which affects the dynamics of the Higgs
field.

Although our numerical lattice simulation is performed based on the
action given in Eq.\ \eqref{einac}, it is also instructive to consider
the frame in which the Higgs field is canonically normalized.  Such a
frame can be realized with the following transformation:
\begin{align}
  h_c \equiv e^{-\chi/2} h.
  \label{hcanonical}
\end{align}
Then, the total action is found to be
\begin{align}
  S=\int d^4x \sqrt{-g}
  \left[-\frac{M_{\rm Pl}^2}{2}R
    +\frac{1}{2}g^{\mu\nu}\partial_\mu\phi\partial_\nu\phi
    +\frac{1}{2}g^{\mu\nu}\partial_\mu h_c \partial_\nu h_c
    - \tilde{V} (\phi, h_c)\right],
\end{align}
where\footnote
{As shown in Eq.\ \eqref{Vtilde}, the Higgs quartic coupling we
  observe should be $\lambda+\frac{3\mu^2}{M_{\rm Pl}^2}\xi^2$.  With
  the model parameters of our interest, i.e., $\mu\simeq 3.1\times
  10^{13}\ {\rm GeV}$ and $\xi\lesssim O(1)$, the second term is
  numerically irrelevant and we neglect its effects.}
\begin{align}
  \tilde{V} (\phi, h_c) =
  \frac{3\mu^2M_{\rm Pl}^2}{4}
  (1-e^{-\chi})^2
  + \frac{1}{2} m_{\rm eff}^2 h_c^2
  + \frac{1}{4} \left( \lambda + \frac{3\mu^2}{M_{\rm Pl}^2} \xi^2 \right)
  h_c^4,
  \label{Vtilde}
\end{align}
with 
\begin{align}
  m_{\rm eff}^2 \equiv 
  \frac{1}{2 \sqrt{-g}}
  \partial_\mu (\sqrt{-g} g^{\mu\nu} \partial_\nu \chi)
  - \frac{1}{4} g^{\mu\nu} (\partial_\mu \chi) (\partial_\nu \chi)
  + 3\mu^2 (1- e^{-\chi}) \xi.
\end{align}
Because the inflaton field is (almost) homogeneous, $m_{\rm eff}^2$
can be expressed as
\begin{align}
  m_{\rm eff}^2 &\simeq
  \frac{1}{\sqrt{6}M_{\rm Pl}} (\ddot{\phi} + 3 H \dot{\phi})
  - \frac{1}{6 M_{\rm Pl}^2} \dot{\phi}^2
  + 3\mu^2
  \left[ 
    1 - \exp \left( -\sqrt{\frac{2}{3}}\frac{\phi}{M_{\rm Pl}} \right)
    \right]
    \xi ,
    \label{meff2}
\end{align}
where $H$ is the expansion rate of the universe.
Using the relation $R\simeq(\dot\phi^2-4V)/M_{\rm Pl}^2$, 
  $m_{\rm eff}^2$ can be also expressed as
\begin{align}
  m_{\rm eff}^2 \simeq
  -\xi R + \left(\xi-\frac{1}{6}\right)\left(\frac{\dot\phi^2}{M_{\rm Pl}^2}
  + \sqrt{6}\frac{\partial_\phi V}{M_{\rm Pl}}\right).
  \label{meff2-2}
\end{align}
We can see that the effective Higgs mass squared is dependent on the
inflaton amplitude.  During the inflation, the inflaton is slowly
rolling with its amplitude much larger than $M_{\rm Pl}$ and hence, if
$\xi$ is sizable, $m_{\rm eff}^2\simeq 3\xi\mu^2 \simeq 12\xi H_{\rm
  inf}^2$ during the inflation.  In particular, if $\xi\gtrsim
O(0.1)$, $m_{\rm eff}$ becomes of the order of the expansion rate and
the quantum fluctuation during the inflation is suppressed.  On the
contrary, in the preheating epoch, $\phi$ is oscillating and hence
$m_{\rm eff}^2$ becomes highly time-dependent.  It is also clearly
seen in Eq.\ (\ref{meff2-2}) that there appear additional terms
proportional to $(\xi-1/6)$ which is characteristic for the
Starobinsky inflation model.  Thus we cannot simply apply the bound on
$\xi$ obtained for inflation models with Einstein gravity in the case
of Starobinsky inflation.  In the following sections, we will see
details of the Higgs dynamics with fully taking account of these
effects.

\section{Instability of the Electroweak Vacuum}
\label{sec:instability}
\setcounter{equation}{0}

Now, we are in the position to discuss the stability of the
electroweak vacuum in the Starobinsky inflation model.  In this
section, we give an overview of the Higgs dynamics.  A detailed study of
the Higgs dynamics based on a lattice simulation will be given in the
next section.  In the Starobinsky inflation model, the vacuum
instability may be a serious issue in two epochs: inflationary epoch
and preheating epoch.  The Higgs dynamics in these epochs are
considered in the following, taking into account important features in
the Starobinsky model.

\subsection{Higgs instability in the early Universe: $|\xi|\ll 1$}

During the inflationary epoch, the Higgs field acquires quantum
fluctuation.  In particular, if the effective mass of the Higgs during
inflation is much smaller than the expansion rate $H_{\rm inf}$, the
amplitude of the quantum fluctuation is typically $H_{\rm inf}$.  In
the case of the Starobinsky inflation, such a quantum fluctuation is
dangerous because $H_{\rm inf}\simeq 1.6\times 10^{13}\, {\rm GeV}$ is
much larger than the instability scale $\Lambda_I$.  In particular,
when $|\xi|\ll 1$, for which the effective Higgs mass during the
inflation is negligible, the Higgs amplitude becomes as large as
$H_{\rm inf}$ within $O(10)$ $e$-folds even if the initial amplitude
vanishes~\cite{Kobakhidze:2013tn,Fairbairn:2014zia,Hook:2014uia,Herranen:2014cua,Kamada:2014ufa,Kearney:2015vba,Espinosa:2015qea,Mantziris:2020rzh}.\footnote
{If the Higgs amplitude is much larger than $H_{\rm
    inf}/\sqrt{|\lambda|}$ at the horizon exit, the Higgs field may
  roll to the true vacuum during the inflation. The inflation is then
  terminated due to the negatively large vacuum energy of Higgs
  potential. We do not consider such a case.}

If the Higgs amplitude $h$ becomes larger than $\sim \Lambda_I$ during
inflation due to the quantum fluctuation, $h$ may have run-away
behavior because of the negative quartic coupling for $h\gtrsim
\Lambda_I$, resulting in a failure to realize the electroweak vacuum
after inflation.  The detailed evolution of the Higgs amplitude is
model dependent; in the case of our interest, the evolution of the
Higgs amplitude should be studied including the effects of
Higgs-inflaton coupling.  In particular, in the case of the
Starobinsky inflation, the effective mass of the Higgs is induced, as
shown in the previous section, which may affect the dynamics of the
Higgs field.

In order to see how the Higgs field evolves if the initial amplitude
is as large as $\sim H_{\rm inf}$, we solve the classical equation of
motion.  Here, we neglect the spatial dependence of the Higgs field
because the non-vanishing Higgs amplitude due to the quantum
fluctuation is particularly important for the over-horizon mode.  In
addition, we consider the case that the energy density of the Higgs is
sub-dominant.  Then, in the frame in which the Higgs field is canonically
normalized, the evolution equation is given by
\begin{align}
  \ddot{h}_c + 3 H \dot{h}_c + 
  m_{\rm eff}^2 h_c + \lambda (Q=h_c) h_c^3 = 0,
  \label{EoM:h-zeromode}
\end{align}
where, in the present calculation, the expansion rate is evaluated as
\begin{align}
  H = 
  \sqrt{
    \frac{1}{3M_{\rm Pl}^2}
  \left[
    \frac{1}{2} \dot{\phi}^2 + \frac{1}{2} \dot{h}_c^2
    + \tilde{V} (\phi, h_c)
    \right]
  }.
\end{align}
We numerically solve the above differential equation and the equation
of motion (EoM) of the inflaton simultaneously.  The initial condition
is imposed at the end of the inflation (see Eq.\ \eqref{t*}).

We first consider the case of $|\xi|\ll 0.1$, for which the effective
mass during the inflation is negligible (see the discussion in the
previous Section).  Then, the Higgs amplitude at the end of inflation
is expected to be $\sim H_{\rm inf}$ or larger.  We numerically solve
Eq.\ \eqref{EoM:h-zeromode} with such an initial condition to see if
the electroweak vacuum can be realized in the present epoch.

\begin{figure}[t]
  \centering
  \includegraphics[width=0.65\textwidth]{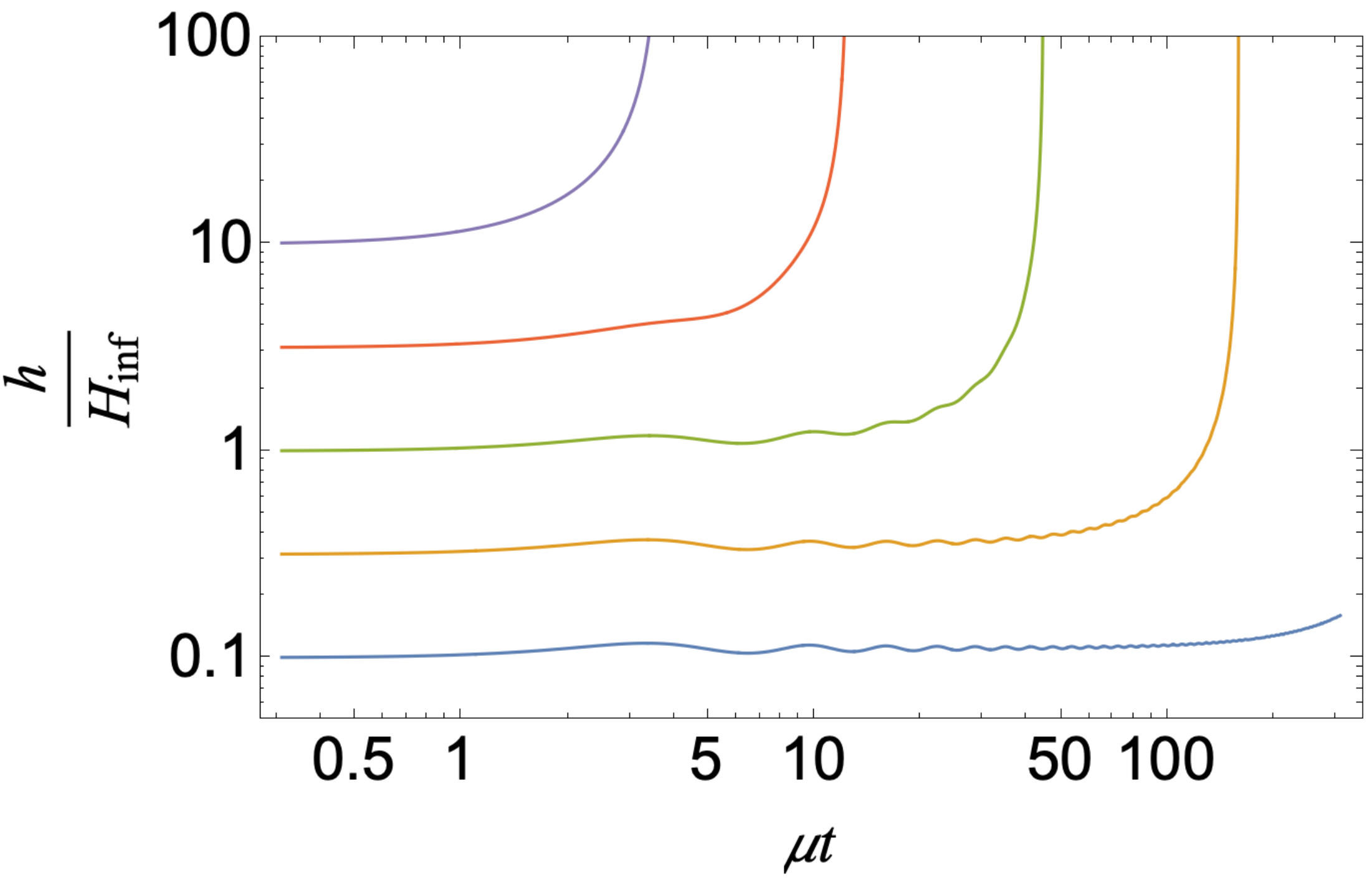}
  \caption{Evolution of the Higgs amplitude after inflation, taking
    $h(t_*)=0.1H_{\rm inf}$ (blue), $0.3H_{\rm inf}$ (orange), $H_{\rm
      inf}$ (green), $3H_{\rm inf}$ (red), and $10H_{\rm inf}$
    (purple), from the bottom to the top.  The non-minimal coupling is
    taken to be $\xi=0$.}
  \label{fig:infstability}
\end{figure}

In Fig.\ \ref{fig:infstability}, we show the evolution of the Higgs
amplitude as a function of time, taking several different values of
$h(t_*)$ and $\xi=0$.  We can see that the Higgs amplitude shows the
run-away behavior when $h(t_*)\gtrsim 0.1 H_{\rm inf}$. 
Our results indicate that, in the Starobinsky inflation model, the
electroweak vacuum at the present universe cannot be realized if
$|\xi|\ll 1$.

The instability due to the quantum fluctuation during the inflation
can be avoided if the non-minimal coupling of the Higgs to gravity is
sizable.  As discussed in the previous section, if $\xi\gtrsim
O(0.1)$, the effective mass of the Higgs during the inflation is as
large as $\sim H_{\rm inf}$, with which the $h(t_*)$ can be much
smaller than $H_{\rm inf}$.  It is known that the quantum fluctuation
during the inflation is suppressed exponentially if $\xi>
\frac{3}{16}$ \cite{Espinosa:2015qea}; in the following discussion, we
consider such a case.

\subsection{Higgs instability during the preheating}

The quantum fluctuation of the Higgs field during the inflation can be
suppressed if $\xi\gtrsim O(0.1)$.  In such a case, however, the Higgs
amplitude may be amplified due to the parametric or tachyonic resonance at the
preheating after the inflation~\cite{Herranen:2015ima,Ema:2016kpf}.

Importance of the parametric resonance can be understood by studying
the behavior of the Higgs effective mass in the preheating epoch.
After the inflation, the inflaton starts to oscillate around the
minimum of the potential with the amplitude smaller than the Planck
scale.  Then the effective mass of the Higgs (\ref{meff2}) in the preheating epoch
is approximately given by
\begin{align}
  \left. m_{\rm eff}^2 \right|_{\rm preheating} \simeq
   (6\xi - 1)\frac{\mu^2 \phi}{\sqrt{6} M_{\rm Pl}}
  +(1-2\xi)\frac{\mu^2 \phi^2}{2M_{\rm Pl}^2}-\frac{\dot\phi^2}{6M_{\rm Pl}^2}.
  \label{meff2(preheating)}
\end{align}
The first term is dominant for $|\phi|\ll M_{\rm Pl}$ and we focus on
it for the moment.\footnote
{ Note, however, that it is the second and
  third terms that give non-vanishing contributions to the effective mass
  squared after time average. By using $\langle\dot\phi^2\rangle\simeq
  \mu^2\langle\phi^2\rangle$, we find $\langle m_{\rm eff}^2
  |_{\rm preheating}\rangle \simeq (1-3\xi) H^2$. Thus it gives
  tachyonic mass for $\xi > 1/3$.  }
With this oscillating
effective mass, the evolution of the Higgs amplitude is well described
by the Mathieu equation.\footnote{ In the limit of small inflaton
  oscillation amplitude, the first term of (\ref{meff2(preheating)})
  describes the perturbative decay of the inflaton into the Higgs
  boson pair (c.g. Refs.\cite{Moroi:2020has,Moroi:2020bkq}).  }  For
the time scale much shorter than $H^{-1}$, for which we can neglect
the effect of the cosmic expansion and approximate the motion of the
inflaton as
\begin{align}
  \left. \phi \right|_{\rm preheating} 
  \sim \bar{\phi} \cos \mu t,
\end{align}
the Fourier amplitude of the Higgs, denoted as
$h_k$ with $k$ being the wave number, is governed by
\begin{align}
  \left(\frac{d^2}{dz^2} +A_k+2q \cos 2z \right)h_k=0,
  \label{Mathieu}
\end{align}
where $z=\mu t/2$, $A_k=4k^2/\mu^2$ and
\begin{align}
  q \equiv \sqrt{\frac{2}{3}} \frac{\bar{\phi}}{M_{\rm Pl}} (6\xi-1).
  \label{q} 
\end{align}
Here, the effect of the quartic coupling, which is unimportant unless
the Higgs amplitude becomes large, is neglected.  At the onset of the
preheating epoch, at which the inflaton amplitude is order of
magnitude smaller than $\sim M_{\rm Pl}$, the broad resonance
condition, $q\gtrsim 1$, is satisfied for $\xi\gtrsim O(1)$.  When
$\xi$ is larger than a few, the Higgs Fourier amplitudes in the
resonance modes become significantly populated.  Such a tachyonic
preheating process may make the Higgs amplitude larger than
$\Lambda_I$ and cause a run-away behavior of the Higgs field.  Thus,
we expect that the non-minimal coupling constant $\xi$ is bounded from
above to realize the electroweak vacuum in the present universe.

In deriving the upper bound on $\xi$, a careful analysis is necessary.
Once the Higgs amplitude becomes sizable, the quartic interaction of
the Higgs becomes important.  In addition, as discussed in the
previous section, the EoMs of the inflaton and Higgs are coupled so
that the EoMs should be solved simultaneously to take account of the
effects of the back reaction to the inflaton dynamics from the
particle creation due to the parametric resonance.  For the precise
study of the dynamics of the inflaton and Higgs fields taking into
account the above mentioned effects as well as the cosmic expansion,
we perform a numerical lattice simulation in the next Section.

\section{Higgs Dynamics after Inflation}
\label{sec:lattice}
\setcounter{equation}{0}

In this section, we study the dynamics of the Higgs field in the
preheating epoch in detail.  Even if the Higgs quantum fluctuation
during the inflation is suppressed by, for example, the mass term from
the Higgs non-minimal coupling, the Higgs field may be resonantly
excited in the preheating epoch.  Once the averaged amplitude of the
Higgs field becomes larger than $\sim\Lambda_I$, Higgs may show the
run-way behavior because of the negative quartic coupling
\cite{Ema:2016kpf, Ema:2017loe, Enqvist:2016mqj}.  As mentioned
earlier, the effect of the parametric resonance is expected to be more
important for larger value of $\xi$.  Too large $\xi$ should result
in the instability of the electroweak vacuum. The upper bound on $\xi$
is studied in detail in the following.

The resonant production is effective, particularly for the modes in the
instability bands.  Because the oscillation frequency of the inflaton
is $\sim\mu$ in the preheating epoch, the wave number of the
instability modes are $k\sim\frac{1}{2}\mu$, $\mu$, $\frac{3}{2}\mu$,
$\cdots$.  For the study of the parametric resonance, the inclusion of the
spatial dependence of the Higgs amplitude is crucial.  We should also
consider the effects of cosmic expansion.  In order to take
account of these, we use a numerical lattice simulation to study the
dynamics of the Higgs field after the Starobinsky inflation.

We perform our lattice simulation based on the Einstein frame action
given in Eq. \eqref{einac}.  The equation of motion of the inflaton is
given by
\begin{align}
  \ddot{\phi} + 3H\dot{\phi} - \frac{1}{a^2}\partial^2_i\phi
  +\frac{1}{\sqrt{6}M_p}e^{-\chi}
  \left[ \dot{h}^2-\frac{1}{a^2}(\partial_ih)^2\right]
  +\frac{\partial V}{\partial \phi}=0,
  \label{ddotphi}
\end{align}
while that of the Higgs field is
\begin{align}
  \ddot{h} + 3H\dot{h} - \frac{1}{a^2}\partial^2_ih 
  - \sqrt{\frac{2}{3}}\frac{1}{M_p}
  \left[\dot{\phi}\dot{h} -\frac{1}{a^2}(\partial_i\phi)(\partial_ih)\right]
 + e^\chi\frac{\partial V}{\partial h}=0.
  \label{ddoth}
\end{align}
We evaluate the expansion rate by using the spatially averaged energy
density as
\begin{align}   
   H = \sqrt{ \frac{\langle \rho\rangle}{3M_{\rm Pl}^2} },
\end{align}
where
\begin{align}
  \rho = \frac{1}{2}\left[\dot{\phi}^2+\frac{1}{a^2}(\partial_i\phi)^2\right]
  +\frac{1}{2}e^{-\chi}\left[\dot{h}^2+\frac{1}{a^2}(\partial_ih)^2\right]
  +V(\phi,h).
\end{align}
Compared to the conventional inflation models without the $R^2$ term,
there are several differences in the equations of motion; in
Eq.\ \eqref{ddoth}, we can find cross terms of the inflaton and the
Higgs in the square bracket and also a factor of $e^\chi$ in front of
the derivative of the potential, which do not exist in the case of the
conventional inflation.  They may affect the dynamics of the Higgs
field.

We are interested in the Higgs dynamics with the scalar potential
given in Eq.\ \eqref{potential}.  The potential is, however, unbounded
below with taking into account the scale dependence of the quartic
coupling constant $\lambda$.  Such a potential is problematic for our
lattice simulation because it makes the numerical calculation
unstable.  We add a $h^6$ term to stabilize the potential to cure this
difficulty.  In our lattice simulation, we use the following
potential:
\begin{align}
  V(\phi,h) = 
  \frac{3\mu^2M_{\rm Pl}^2}{4}
  \left(1-e^{-\chi}+\frac{\xi}{M_{\rm Pl}^2}e^{-\chi}h^2\right)^2
  + \frac{\lambda (Q=\sqrt{ \langle h^2 \rangle })}{4} e^{-2\chi} h^4
  + \frac{c}{6 M_{\rm Pl}^2} e^{-2\chi}h^6,
  \label{Vwithh^6}
\end{align}
where $c$ is a positive constant.  With our choice of $c$, the Higgs
potential at $H\sim\Lambda_I$ is almost unaffected although the $h^6$
term changes the behavior of the potential at large Higgs amplitude.
Thus, the onset of the instability is not affected by the $h^6$ term
in our analysis, as we show in the following.  The Higgs potential
given in Eq.\ \eqref{Vwithh^6} has its minimum $h=h_{\rm min}$ which
is given by
\begin{align}
  h_{\rm min} \simeq \sqrt{\frac{|\lambda (Q=h_{\rm min})|}{c}} M_{\rm Pl},
\end{align}
where $\lambda (Q=h_{\rm min})<0$ is assumed.  We take $c=2\times
10^{3}$ unless otherwise mentioned.

In the lattice simulation, the field amplitudes of the inflaton and
the Higgs field at the lattice sites are followed by numerically
solving Eqs.\ \eqref{ddotphi} and \eqref{ddoth}.  We modify the GABE
code \cite{Child:2013ria}, which uses the second-order Runge-Kutta
method to solve differential equations, to simulate the inflaton-Higgs
coupled system of our interest.  We start the calculation from the end
of the inflation, i.e., $t=t_*$ (see Eq.\ \eqref{t*}).  We take
the initial box size $L=20/\mu$ with the number of grids $N=128$ per edge.
The time step is taken to be $dt=10^{-3}/\mu$. The scale factor is
normalized as $a (t_*)=1$.  We are paying particular attention to the
Higgs production by the parametric resonance.  Then, we are interested
in the Higgs fluctuations with the wave number of the order of
$\sim\mu$, corresponding to the wavelength of $\sim \frac{2\pi}{\mu}$.
On the contrary, the lattice spacing is $\sim 0.2 a(t) \mu^{-1}$ with
our choice of the lattice parameters.  Then, the lattice spacing may
become too large to resolve the Higgs fluctuation from the parametric
resonance when $a(t)\gtrsim 10$ or so, which is the case when $\mu
t\gtrsim O(10)$.  (We found that the scale factor is $5.0$, $5.7$, and
$6.4$ for $t=20\mu^{-1}$, $25\mu^{-1}$, and $30\mu^{-1}$,
respectively.) We expect that our numerical calculation is reliable as
far as $\mu t\lesssim O(10)$.

For the initial field values of the inflaton and canonically
normalized Higgs, we presume that they originate from the quantum
fluctuations at $t=t_*$; we firstly evaluate them with neglecting the
cosmic expansion. In the lattice simulation, we study the evolutions of
the field values at lattice sites $\vec{x}=\frac{L}{N}(n_x,n_y,n_z)$
with $0\leq n_{x,y,z}<N$.  The field operators at the lattice sites
can be expressed as
\begin{align}  
  \hat{X}^{\rm (lattice)} (\vec{x},t) = 
  \frac{1}{L^{3/2}} \sum_{\vec{k}}
  \frac{1}{\sqrt{2\omega_k}}
  \left[ \hat{a}_{\vec{k}} (t) e^{i\vec{k}\vec{x}} + {\rm h.c.} \right],
\end{align}
where $X=\phi$ and $h_c$ (see Eq.\ \eqref{hcanonical}). Here,
$\vec{k}=\frac{2\pi}{L}(n'_x,n'_y,n'_z)$ with $0\leq n'_{x,y,z}<N$ and
$\omega_k^2=\vec{k}^2+\partial^2 \tilde{V}/\partial X^2$.  Then, after
the conventional canonical quantization, we find
$[\hat{a}_{\vec{k}},\hat{a}^\dagger_{\vec{k}'}]=\delta_{\vec{k},\vec{k}'}$
and $\langle \hat{a}_{\vec{k}} \hat{a}_{\vec{k}}^\dagger\rangle=1$ (with
$\langle\cdots\rangle$ being the vacuum expectation value).  We set
the initial values of the scalar amplitudes as
\begin{align}  
  X ^{\rm (lattice)} (\vec{x},t_*) = 
  \frac{1}{L^{3/2}} \sum_{\vec{k}}
  \frac{1}{\sqrt{2\omega_k}}
  \left( \tilde{X}_{\vec{k}} e^{i\vec{k}\vec{x}} + {\rm c.c.} \right),
\end{align}
where $\tilde{X}_{\vec{k}}$'s are regarded as statistical variables.
The statistical properties of $\tilde{X}_{\vec{k}}$'s are determined
so that $\langle \hat{X}^{\rm (lattice)} (\vec{x},t_*)\hat{X}^{\rm
  (lattice)} (\vec{x}',t_*)\rangle = \langle {X}^{\rm (lattice)}
(\vec{x},t_*){X}^{\rm (lattice)} (\vec{x}',t_*)\rangle_{\rm stat}$
(with $\langle\cdots\rangle_{\rm stat}$ being statistical average).
Then, we find
$\langle\tilde{X}_{\vec{k}}^*\tilde{X}_{\vec{k}}\rangle_{\rm
  stat}=\frac{1}{2}$; in the lattice simulation,
$\tilde{X}_{\vec{k}}$'s are sampled by assuming that
$\mbox{Re}\tilde{X}_{\vec{k}}$ and $\mbox{Im}\tilde{X}_{\vec{k}}$ obey
Gaussian distribution $N(0,\frac{1}{4})$.  The time dependence of
$\tilde{X}_{\vec{k}}$ is given by $\tilde{X}_{\vec{k}}\propto
a^{-1}e^{-i\omega_k t}$, and the initial time derivative is
\begin{equation}
	\dot{\tilde{X}}_{\vec{k}}=(- i\omega_k-H(t_*))\tilde{X}_{\vec{k}}.
\end{equation}   
Finally, fluctuations of the canonically normalized Higgs field and
its time derivative are rescaled back to those of the original Higgs
field and then added to the homogeneous parts.

In order to check the reliability of our numerical analysis, we have
performed the analysis with $N=256$ (as well as $N=128$) for $\xi=1.4$
and $1.8$ taking central values of the standard-model couplings.  For $\xi=1.4$,
we found that the electroweak vacuum is stable until $\mu t\lesssim
30$ for both choices of the number of grids while two results show
difference at a later epoch; at $\mu t\gtrsim 30$, there is no sign of
the instability for $N=128$ while, for $N=256$, the Higgs variance
shows a significant increase. In addition, for $N=256$, we found that
the detail of the behavior at $\mu t\gtrsim 30$ is dependent on the
initial configuration.  As we have mentioned earlier, the lattice
spacing becomes of the same order as the wavelength of our interest
when $\mu t\sim O(10)$, which may be the cause of the difference.  For
$\xi=1.8$, destabilization happens at $\mu t\sim 20$ both for $N=128$
and $256$ and two choices of the number of grids does not show the
qualitative difference.\footnote
{We note that $N$ cannot be taken too large because, in the present
  prescription, the dispersion relation of the Higgs may be
  significantly altered by the initial fluctuation.  Rigorously
  speaking, a renormalization is necessary to subtract such a
  correction (c.f., Ref.~\cite{Ema:2016kpf}).  In the present case,
  however, it is neglected because the effect is unimportant for the
  study of the parametric resonance.  Substituting the initial
  fluctuation into $\lambda\langle h^2\rangle$, the correction to the
  Higgs mass squared is estimated to be $\sim \frac{\lambda}{16\pi^2}
  \frac{(2\pi)^2 N^2}{L^2}$.  It is smaller than the typical momentum
  squared relevant for the parametric resonance (i.e., $\mu^2$) as far
  as $N\lesssim 400 \sqrt{\frac{0.01}{|\lambda|}}$.  In our
  calculation, this condition is met.}
Thus, we expect that our numerical
calculation with $N=128$ is reliable for $\mu t\lesssim 30$ while the
results for $\mu t\gtrsim 30$ may be affected by numerical artifacts. 
In the following, we rely on the numerical results for $\mu t\leq 25$
with taking $N=128$ to derive a bound on $\xi$; we could not increase
$N$ because of the limitation of the computational resource.

\begin{figure}[t]
  \centering
  \includegraphics[width=8cm]{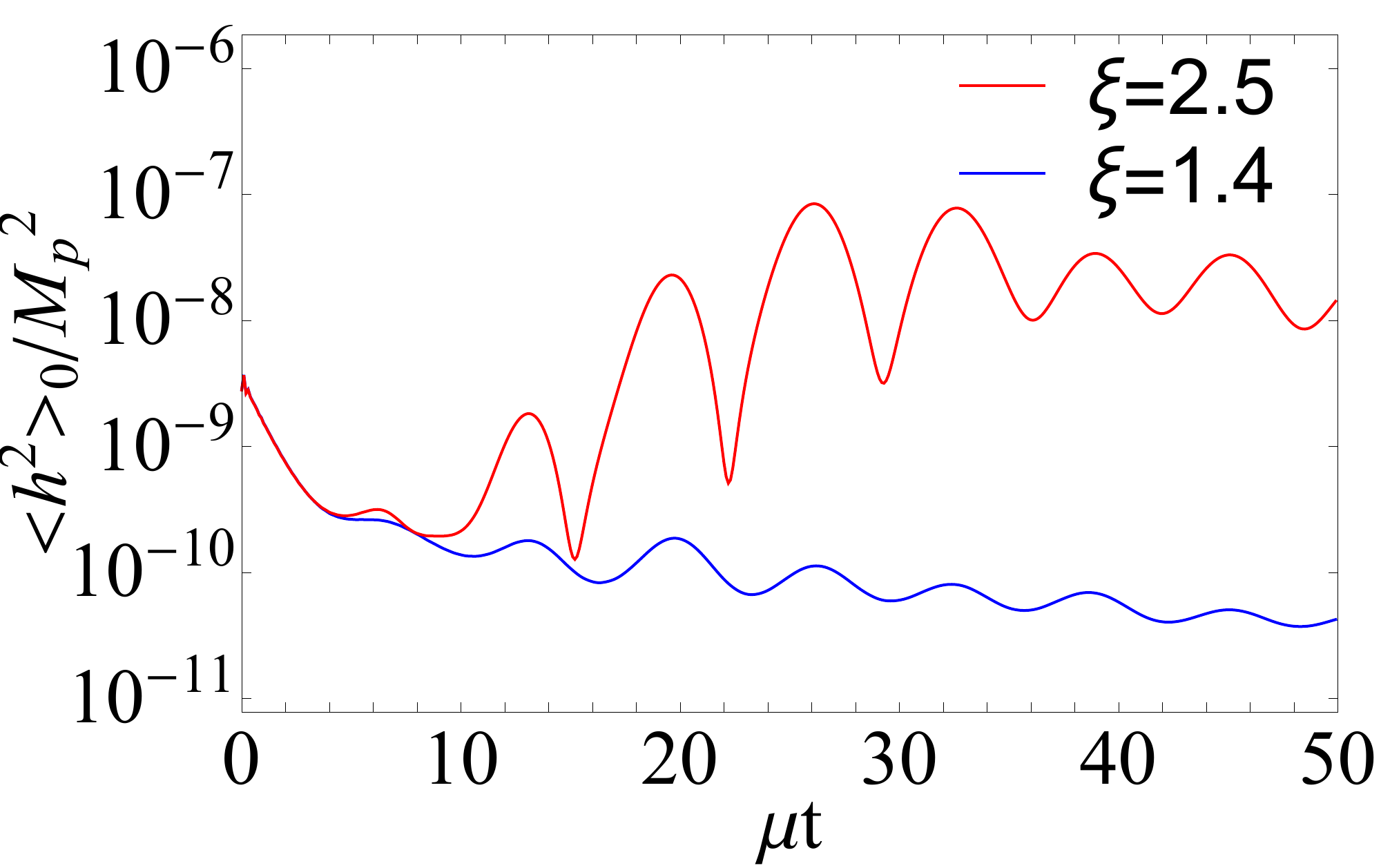}
  \caption{Time evolution of the Higgs field variance $\langle
    h^2\rangle_0$ for $\lambda=0$. The non-minimal Higgs-gravity
    couplings are taken to be $\xi=1.4$ (blue) and $\xi=2.5$ (red). We
    take $\mu t_{\rm{end}}=20$ and $25$ to derive the upper bound on
    $\xi$.}
  \label{fig:resonance}
\end{figure}

If the destabilization happens, the Higgs variance starts to blow up.
The destabilization process may be affected by the scattering (and
thermalization) processes of the Higgs field.  At the epoch of our
interest, i.e., $\mu t\lesssim 30$, we presume that the effects of the
scatterings are not important. During such an epoch, the Higgs
occupation number $\eta$ exponentially increases, while the scattering
cross section is $\sigma\sim \frac{g^4}{4\pi}\frac{1}{\mu^2}$ (with
$g$ being the gauge coupling constant).\footnote{We do not consider the possibly fast decay process, e.g., $H\to t\bar t$, which may be kinematically blocked due to the plasma mass induced by the large Higgs occupation number.   }  Then, the scattering rate is
$\Gamma_{\rm scatt}\sim\eta\mu^3\sigma\sim\frac{g^4}{4\pi}\eta\mu$.
Because the resonance parameter $q$ is at most $\sim 1$ for the case
we consider (see Eq.\ \eqref{q}) and the redshift effect takes the
enhanced modes away from the resonance band, $\eta$ is not expected to
be extremely large and the scattering rate is expected to be smaller
than the expansion rate, which is $O(0.1)\mu$ for $\mu t\lesssim 30$.
Thus, we neglect the effects of the scattering processes.

In the following, we derive a conservative bound on $\xi$
concentrating on the resonance regime.  The time of the end of the
resonance regime, denoted as $t_{\rm{end}}$, is estimated by studying
the Higgs dynamics with $\lambda=0$; the Higgs variance for
$\lambda=0$ is denoted as $\langle h^2\rangle_0$.  After
$t_{\rm{end}}$, the peak value of $\langle h^2\rangle_0$ is expected
to decrease because the effect of the Hubble friction wins over the
effect of the parametric resonance.  In Fig.\ \ref{fig:resonance}, we
show the evolution of $\langle h^2\rangle_0$, taking $\xi=1.4$ and
$\xi=2.5$. We can see that $\langle h^2\rangle_0$ reaches the highest
peak in the time interval of $20\lesssim \mu t\lesssim 30$; we have
checked that, when $1.4\leq\xi\leq 2.5$, the highest peak is realized
during this period.  For larger $\xi$, the parametric resonance stops
at a later epoch.  In the following, we consider the cases with
$1.4\leq\xi\leq 2.5$ and take $\mu t_{\rm{end}}=20$ and $25$.

\begin{figure}[t]
  \begin{subfigure}{0.5\textwidth}
    \centering
    \includegraphics[width=8cm, height=5cm]{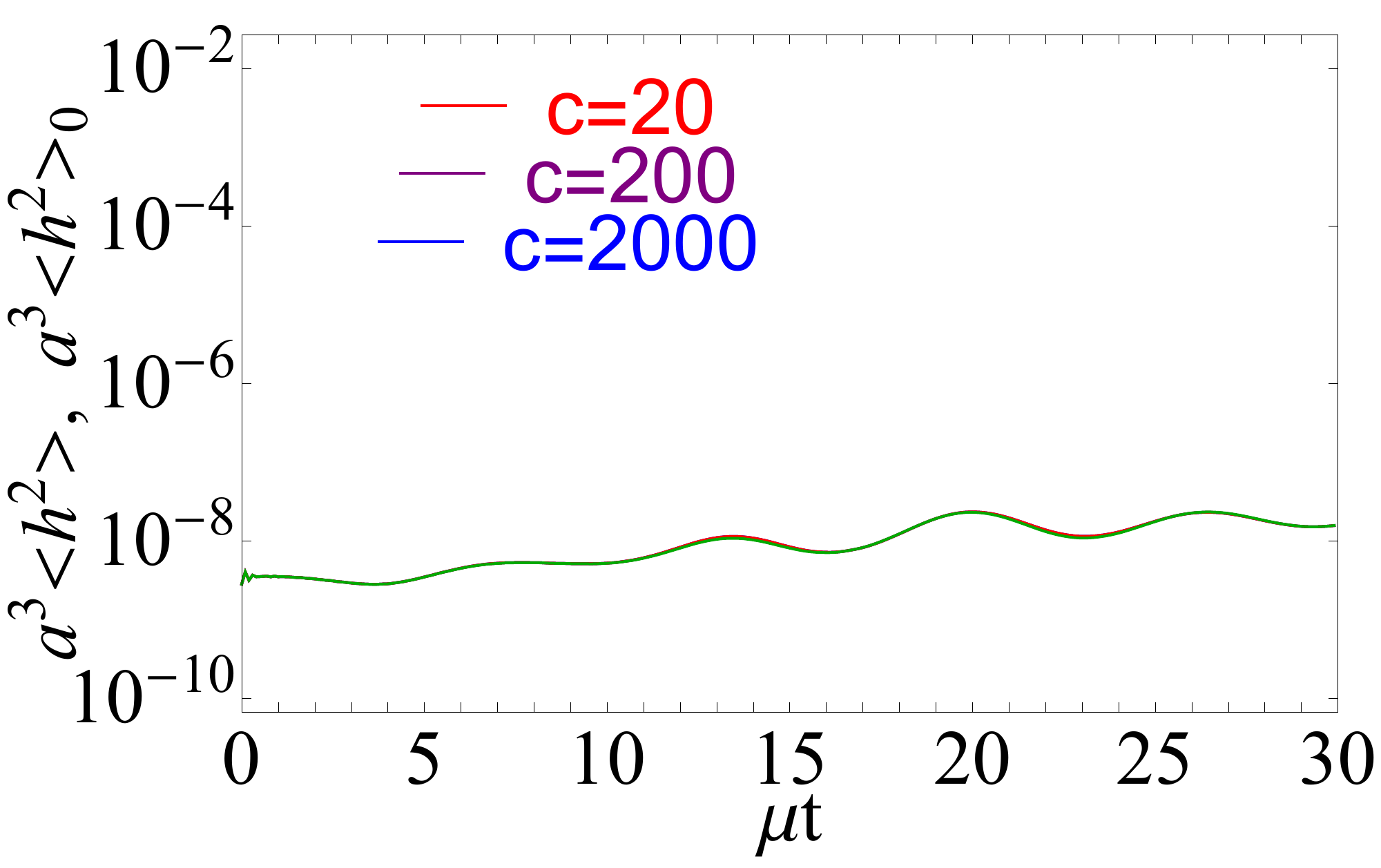}
    \caption{$\xi=1.4$}
  \end{subfigure}
  \begin{subfigure}{0.5\textwidth}
    \centering
    \includegraphics[width=8cm, height=5cm]{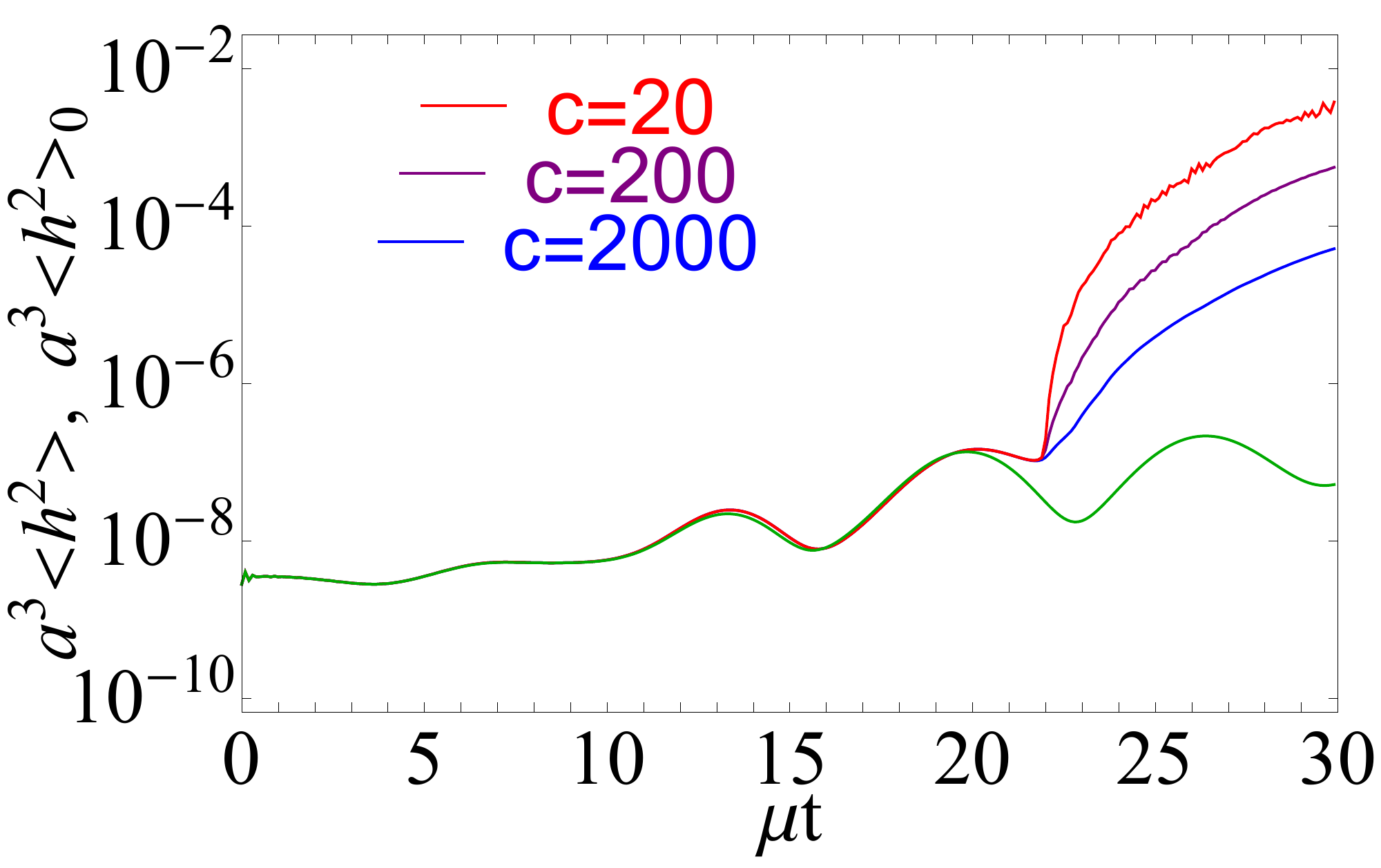}
    \caption{$\xi=1.8$}
  \end{subfigure}
  \caption{Lattice simulation results of the Higgs field variance $\langle h^2\rangle$ for $\xi=1.4$ and $1.8$ and varying values of $c$: 20 (red), 200 (purple) and 2000 (blue). The top quark mass is taken to be $M_t=172$ GeV, and the strong coupling is the central value $\alpha_s=0.1179$. Time evolution of $\langle h^2\rangle_0$ is shown in green line. Both $\langle h^2\rangle$ and $\langle h^2\rangle_0$ are multiplied by $a^3$ and normalized by $M_{\rm Pl}^2$.  }
  \label{fig:cdep}
\end{figure}

In order to quantify the instability of the electroweak vacuum, we use
the fact that the Higgs variance $\langle h^2\rangle$ becomes
significantly larger than $\langle h^2\rangle_0$ once the instability
occurs.  In Fig.\ \ref{fig:cdep}, we show the evolutions of $\langle
h^2\rangle$ and $\langle h^2\rangle_0$ for $\xi=1.4$ and $1.8$; here
we take $c=20$, $200$, and $2000$.   For
the case of $\xi=1.4$, the instability does not occur and $\langle
h^2\rangle$ behaves as $\langle h^2\rangle_0$.  With larger value of
$\xi$, $\langle h^2\rangle$ starts to deviate from $\langle
h^2\rangle_0$ at $t\sim t_{\rm inst}$ and shows significant increase
after $t\sim t_c$; for the case of $\xi=1.8$ shown in
Fig.\ \ref{fig:cdep}, $t_{\rm inst}\simeq 20\mu^{-1}$ and $t_c\simeq
22\mu^{-1}$.  The Higgs variance at $t\lesssim t_c$ is insensitive to
the choice of $c$.  On the contrary, $\langle h^2\rangle$ at $t\gtrsim
t_c$ depends on $c$; we can see that, in such an epoch, $\langle
h^2\rangle$ is approximately proportional to $c^{-1} (t-t_c)^3$.  We
comment that such behavior arises when the universe is filled with
the ``false vacuum region'' with $h\ll h_{\rm min}$ and the ``true
vacuum bubble,'' in which $h\sim h_{\rm min}$, whose wall velocity is
close to the speed of light.\footnote
{We comment that the observation here indicates a new possibility to
  realize a relativistic expansion of bubble walls.  Let us consider a
  scalar field $s$, with its mass smaller than $H_{\rm inf}$, whose
  potential has a negative quartic coupling and a positive
  Planck-suppressed higher dimensional term.  If $s$ has a non-minimal
  coupling $\sim s^2 R$, $s$ is trapped at the origin during the
  inflation, and its amplitude may be parametrically enhanced after
  inflation.  The dynamics of $s$ is similar to that of the Higgs
  studied in our analysis.  If the amplification of the amplitude of
  $s$ is large enough, the tachyonic instability of $s$ may happen,
  resulting in the formation of bubbles in which $s$ is at the minimum
  of its potential. The latent energy carried by $s$ once become the
  kinetic energy of the wall, then transferred to the energy of
  radiation, e.g., with bubble collisions.  Contrary to the case of the
  standard-model Higgs, a viable cosmological scenario is possible
  because we may live in a vacuum with a very large amplitude of $s$.
  Since the phenomena may be similar to that in the strong first-order
  phase transition with relativistic bubble expansion, relevant
  particle production mechanisms may be applicable
  \cite{Falkowski:2012fb,Katz:2016adq, Vanvlasselaer:2020niz,
    Azatov:2021ifm, Azatov:2021irb,Baldes:2021vyz}.  However, the
  gravitational waves due to the bubble wall collisions or sound waves
  may be too high-frequency to be observed in the near future if the
  inflation scale is high.}
The deviation of $\langle h^2\rangle$
from $\langle h^2\rangle_0$ is expected to be a sign of the
instability.  In our analysis, we adopt the following criterion for
the instability:
\begin{align}
  \left. 
  \frac{\langle h^2\rangle-\langle h^2\rangle_0}{\langle h^2\rangle_0}
  \right|_{t=t_{\rm end}}
  > 2.
  \label{instability}
\end{align}
With too large $\xi$, the above condition is met, indicating that the
tachyonic mass induced by Higgs self coupling dominates the total
effective mass and that the instability of the electroweak vacuum is
triggered.

\begin{figure}
  \begin{subfigure}{0.5\textwidth}
    \centering
    \includegraphics[width=8cm, height=5cm]{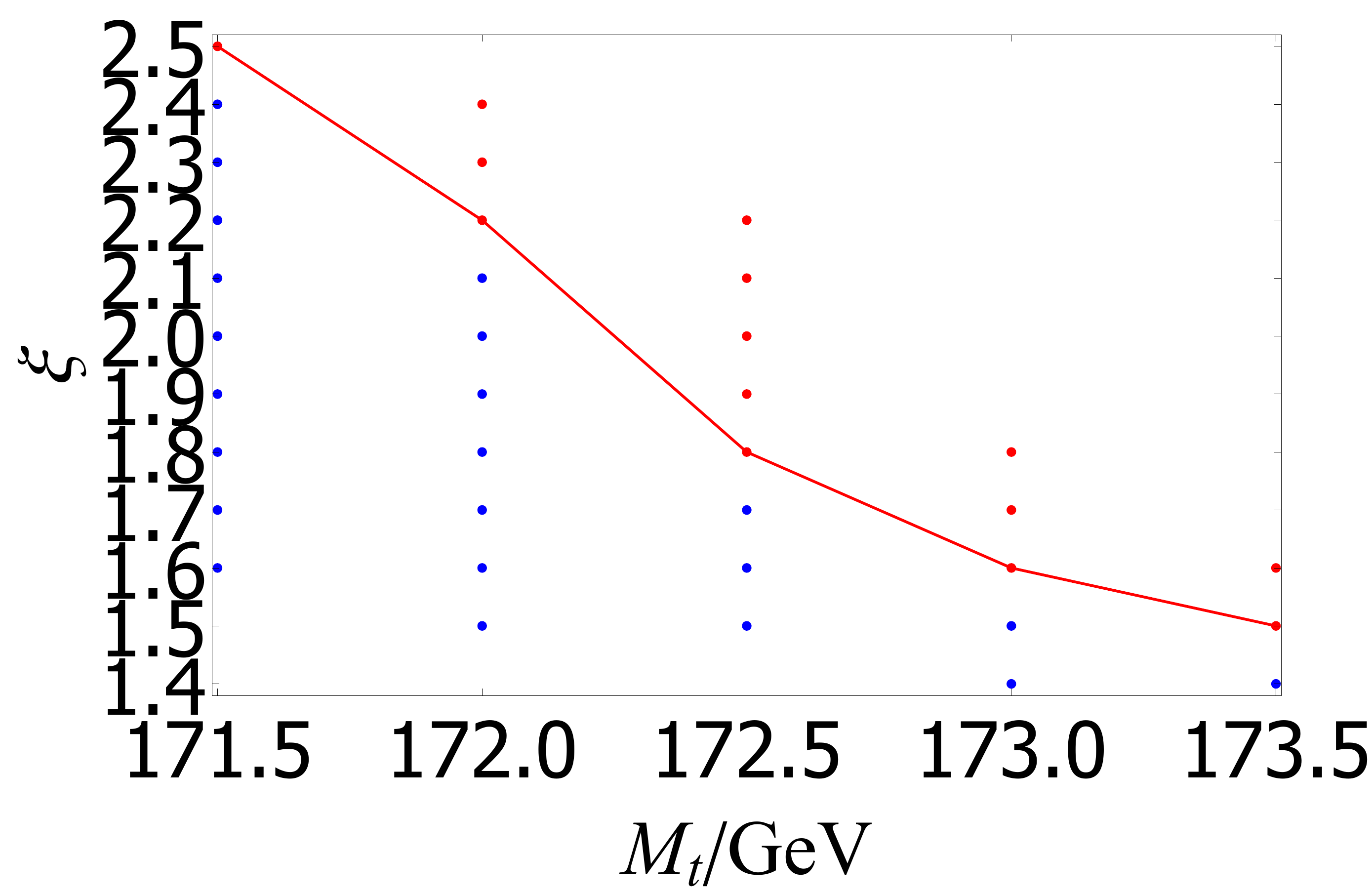}
    \caption{$\mu t_{\rm{end}}=20$}
  \end{subfigure}
  \begin{subfigure}{0.5\textwidth}
    \centering
    \includegraphics[width=8cm, height=5cm]{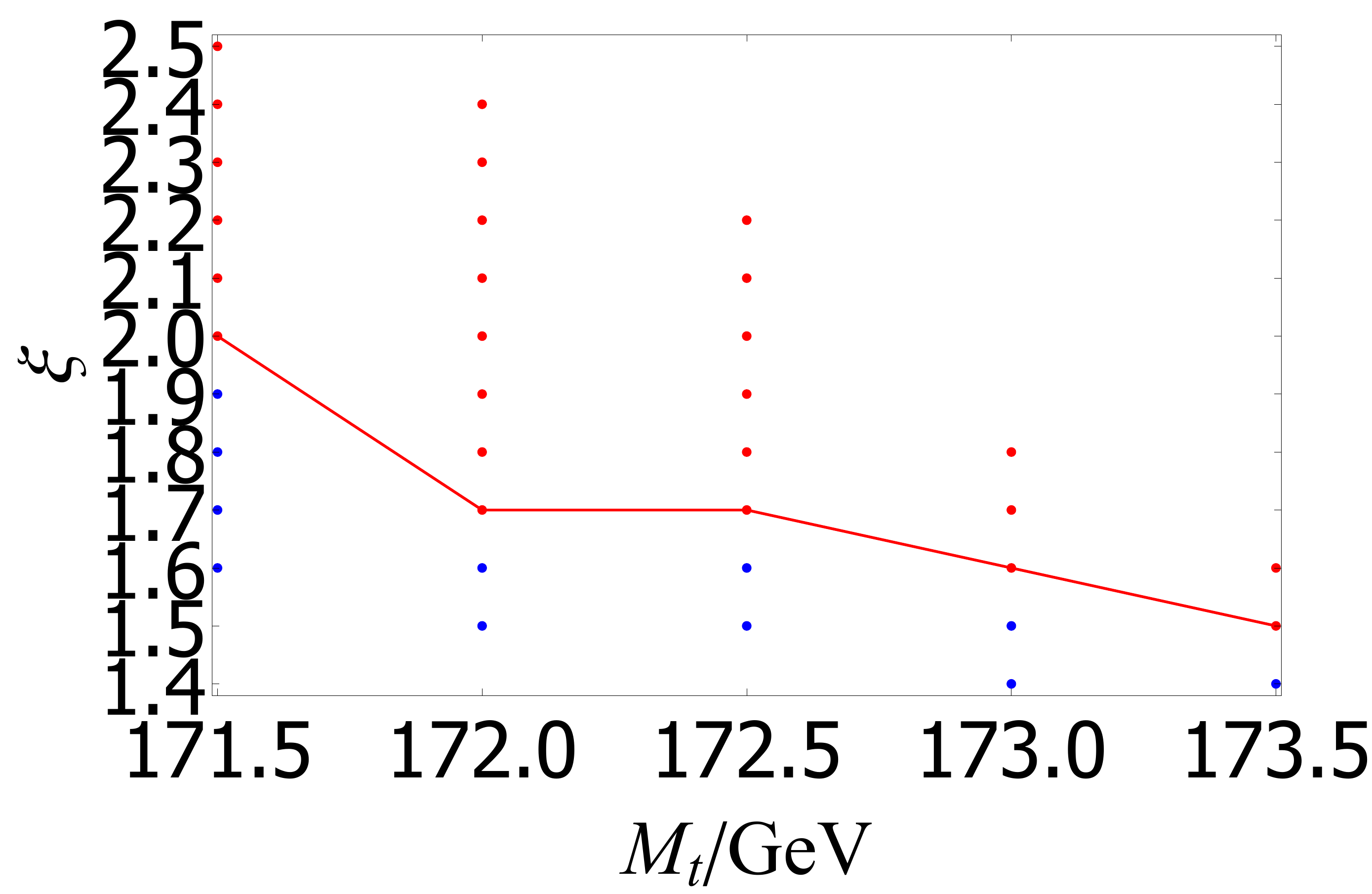}
    \caption{$\mu t_{\rm{end}}=25$}
  \end{subfigure}
  \caption{Vacuum stability bounds on $\xi$ for different values of
    the top quark mass $M_t$, taking $\alpha_s=0.1179$ and $\mu
    t_{end}=20$ and $25$.  The red (blue) dots show the sample
    points with (without) the instability of the electroweak
    vacuum. The red line indicates our upper bounds on $\xi$.
   }
  \label{fig:mtbound}
  \vspace{10mm}
  \begin{subfigure}{0.5\textwidth}
    \centering
    \includegraphics[width=8cm, height=5cm]{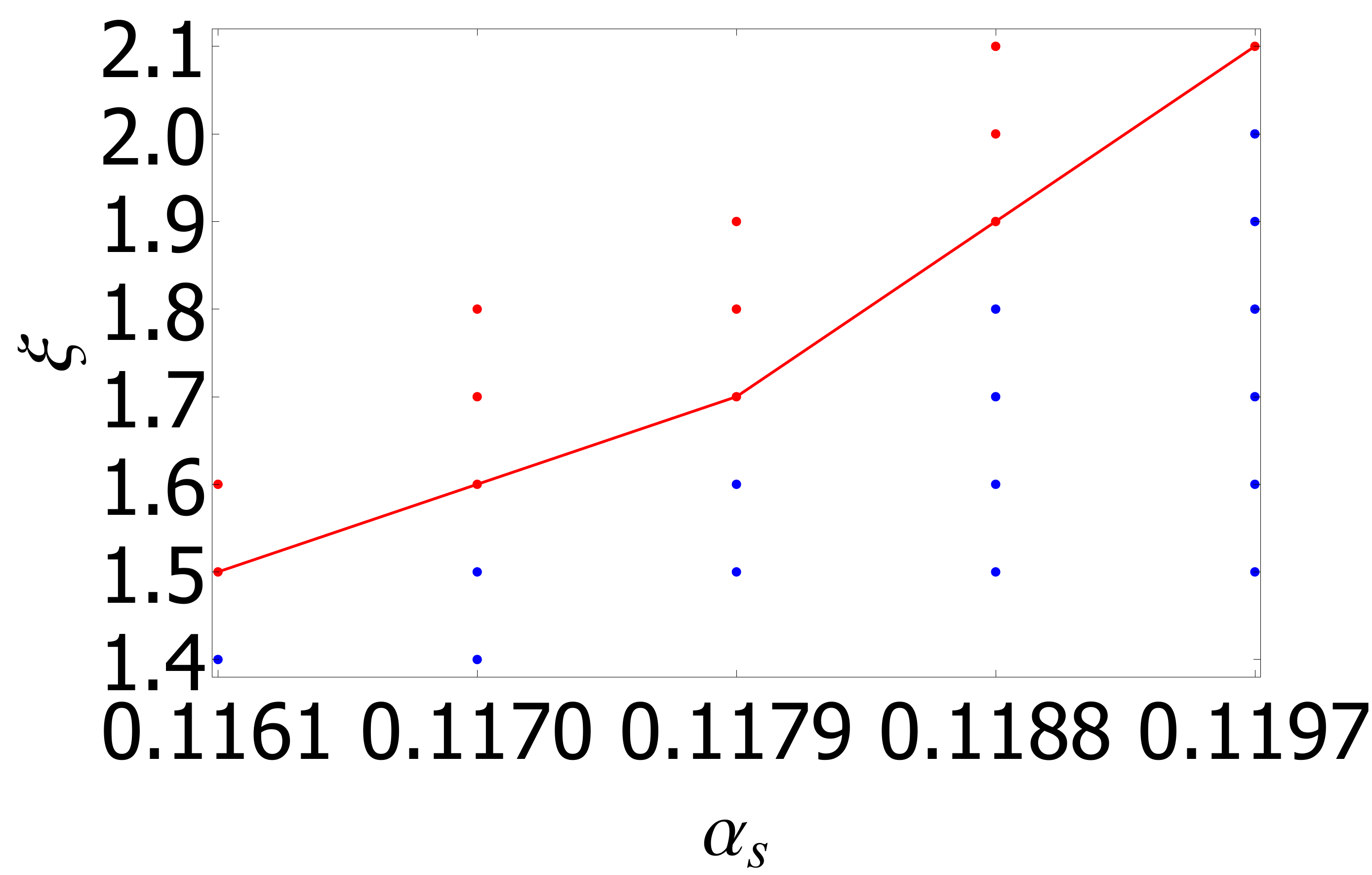}
    \caption{$\mu t_{\rm{end}}=20$}
  \end{subfigure}
  \begin{subfigure}{0.5\textwidth}
    \centering
    \includegraphics[width=8cm, height=5cm]{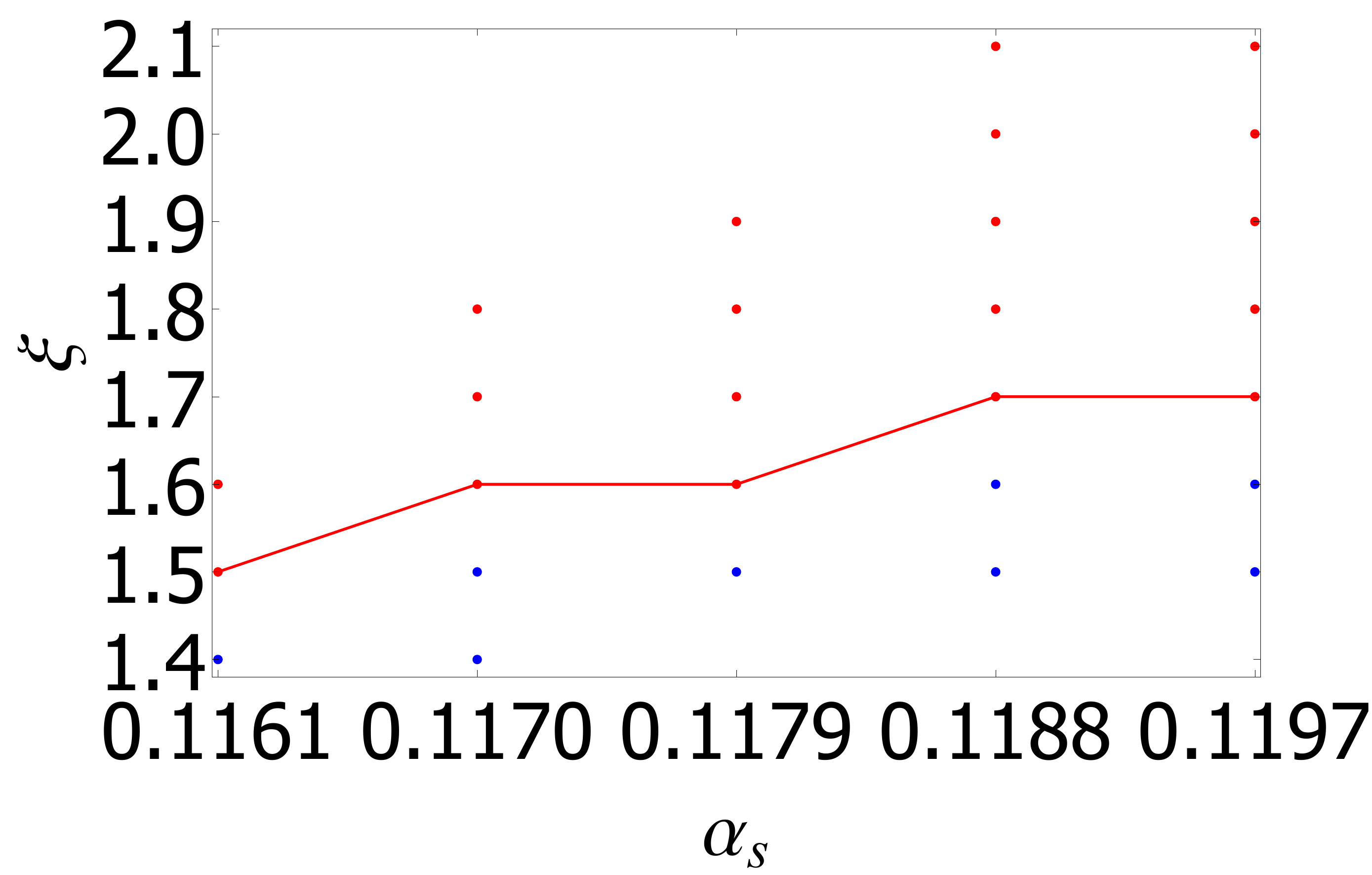}
    \caption{$\mu t_{\rm{end}}=25$}
  \end{subfigure}
  \caption{Vacuum stability bounds on $\xi$ for different values of
    the strong coupling $\alpha_s$, taking $M_t=172.76\ {\rm GeV}$ and
    $\mu t_{\rm end}=20$ and $25$.  The red (blue) dots show the
    sample points with (without) the instability of the electroweak
    vacuum. The red line indicates our upper bounds on $\xi$.}
  \label{fig:alphabound}
\end{figure}

We have studied the behavior of the Higgs variable for several choices
of the non-minimal coupling $\xi$ and the top quark mass $M_t$.  We
take the non-minimal coupling $\xi$ with the interval of $0.1$ and top
quark mass in the range of $171.5-173.5$ GeV with the interval of
$0.5$ GeV.  The sample points on which we perform the lattice
simulation are indicated by the dots on Fig.\ \ref{fig:mtbound}; for
the figure, the central value of the strong coupling constant is used
while $\mu t_{\rm end}$ is taken to be $20$ and $25$.  The red dots on
the figure show the sample points on which the destabilization is
observed (see Eq.\ \eqref{instability}) while the blue ones are sample
points without the sign of instability.  We have connected the red
dots at the boundary, which we regard as an upper bound on the
non-minimal coupling.  We can see that the upper bound on $\xi$
becomes smaller as the top quark becomes heavier.  This is due to the
fact that, with larger top quark mass, $\lambda$ becomes smaller
meaning that the absolute value of the tachyonic mass induced by Higgs
self coupling is more enhanced (see Fig.\ \ref{fig:lambda}).

In order to see how the bound depends on the strong coupling constant,
we also perform the lattice simulation for several values of
$\alpha_s$; the result is shown in Fig. \ref{fig:alphabound} (for
which the top quark mass is taken to be the central value
$M_t=172.76\ {\rm GeV}$).  As in the case of Fig.\ \ref{fig:mtbound},
the red and blue dots indicate the sample points with and without the
sign of the destabilization before $t_{\rm end}$.  We can see that the
upper bound becomes larger for larger value of $\alpha_s$, which is
due to the fact that $\lambda (Q)$ with fixed $Q>M_t$ increases with
the increase of $\alpha_s$.

The upper bound on the non-minimal coupling depends slightly on the
choice of $t_{\rm end}$.  For $20<\mu t_{\rm end}<25$, the upper bound
varies $\sim O(10)\ \%$ and is larger for smaller $t_{\rm end}$.
Adopting the central values of $M_t$ and $\alpha_s$, the upper bound
on $\xi$ is $1.6-1.7$.  The bound is significantly smaller than the
one obtained in the case of conventional inflation models without the
$R^2$ term, which gives $\xi\lesssim 5$ \cite{Ema:2016kpf,Figueroa:2017slm}. 

\section{Conclusions and Discussion}
\label{sec:conclusions}
\setcounter{equation}{0}

We have discussed the stability of the electroweak vacuum during and
after the Starobinsky inflation.  We paid particular attention to the
non-minimal coupling of the Higgs to gravity, and studied the
enhancement of the Higgs amplitude due to the parametric resonance
after the inflation.  Because the Starobinsky inflation requires the
expansion rate during inflation to be larger than the instability
scale of the Higgs potential in the standard model, the quantum
fluctuation during inflation may make the Higgs amplitude larger than
the instability scale, resulting in the run-away behavior of the Higgs
field.  The non-minimal coupling of the Higgs field to the Ricci
scalar is introduced to avoid such instability.  The non-minimal
coupling, however, may induce a parametric-resonance production of the
Higgs after inflation, which may destabilize the Higgs amplitude.  The
effect of the parametric resonance is more enhanced as the non-minimal
coupling constant $\xi$ becomes larger.

We have studied the dynamics of the Higgs field in the preheating
epoch after inflation in detail in the Starobinsky inflation model.
In the case of the Starobinsky inflation, the evolution equation of
the Higgs field differs from that in the case of simple inflation
models (which are based on the Einstein-Hilbert action without the
$R^2$ term).  We used the numerical lattice simulation to follow the
evolution of the Higgs field and investigated the stability of the
Higgs amplitude.  We have seen that the Higgs amplitude is
destabilized if the non-minimal coupling constant $\xi$ is large.
With requiring that the Higgs amplitude does not show the run-away
behavior, we derived an upper bound on the non-minimal coupling
constant $\xi$.  With the central values of the standard-model
parameters, for example, we found that $\xi$ should be smaller than
$\sim 1.6-1.7$ in order to realize the electroweak vacuum in the
present universe.

\vspace{2mm}
\noindent{\it Acknowledgments:} 
This work was supported by JSPS KAKENHI Grant Numbers
16H06490 [TM], 22H01215 [TM, WY], 
17H06359 [KN], 18K03609 [KN],
20H05851 [WY], 21K20364 [WY] and 22K14029 [WY].


\bibliographystyle{jhep}
\bibliography{ref}


\end{document}